\documentclass[aps, prx,twocolumn,longbibliography,superscriptaddress,amsmath,amssymb,floatfix]{revtex4-2}
\usepackage{amssymb}
\usepackage{graphicx}
\usepackage{amsmath}
\usepackage{color}
\usepackage{mathrsfs}
\usepackage{float}
\usepackage{indentfirst}
\usepackage{mathrsfs}
\usepackage{float}
\usepackage{indentfirst}
\usepackage{textcomp}
\usepackage{comment}
\usepackage{mathtools}
\usepackage{natbib,hyperref}
\usepackage{soul}

\begin{document}


\title{Plaquette-type valence bond solid state in the $J_1$-$J_2$ square-lattice Heisenberg model}

\author{Jiale Huang}
\affiliation{Key Laboratory of Artificial Structures and Quantum Control (Ministry of Education),  School of Physics and Astronomy, Shanghai Jiao Tong University, Shanghai 200240, China}

\author{Xiangjian Qian}
\affiliation{Key Laboratory of Artificial Structures and Quantum Control (Ministry of Education),  School of Physics and Astronomy, Shanghai Jiao Tong University, Shanghai 200240, China}

\author{Mingpu Qin} \thanks{qinmingpu@sjtu.edu.cn}
\affiliation{Key Laboratory of Artificial Structures and Quantum Control (Ministry of Education),  School of Physics and Astronomy, Shanghai Jiao Tong University, Shanghai 200240, China}
\affiliation{Hefei National Laboratory, Hefei 230088, China}

\begin{abstract}
We utilize Density Matrix Renormalization Group (DMRG) and Fully Augmented Matrix Product States (FAMPS) methods to investigate the Valence Bond Solid (VBS) phase in the $J_1$-$J_2$ square lattice Heisenberg model. To differentiate between the Columnar Valence Bond Solid (CVBS) and Plaquette Valence Bond Solid (PVBS) phases, we introduce an anisotropy $\Delta_y$ in the nearest neighboring coupling in the $y$-direction, aiming at detecting the possible spontaneous rotational symmetry breaking in the VBS phase. In the calculations, we push the bond dimension to as large as $D = 25000$ in FAMPS, simulating systems at a maximum size of $14 \times 14$. With a careful extrapolation of the truncation errors and appropriate finite-size scaling, followed by finite $\Delta_y$ scaling analysis of the VBS dimer order parameters, we identify the VBS phase as a PVBS type, meaning there is no spontaneous rotational symmetry breaking in the VBS phase. {This study not only supports the presence of PVBS order in the $J_1$-$J_2$ square lattice Heisenberg model}, but also highlights the capabilities of FAMPS in the study of two-dimensional quantum many-body systems.

\end{abstract}

\maketitle

\section{Introduction}
\label{sec:I}

The $J_1$-$J_2$ spin-$1/2$ square lattice Heisenberg model stands as one of the most extensively researched paradigms for studying the frustration effect in quantum many-body systems. 
The competing antiferromagnetic interactions between the nearest and next-nearest neighbors give rise to a rich variety of phases, which makes this model a well-known playground for searching exotic quantum states, such as Quantum Spin Liquid (QSL) and Valence Bond Solid (VBS) \cite{PhysRevLett.63.2148,PhysRevLett.113.027201,PhysRevLett.87.097201,PhysRevLett.66.1773}. 
Understanding these exotic states may be crucial for elucidating the underlying physics of high-temperature superconductors and other strongly correlated materials \cite{science.235.4793.1196,PhysRevB.39.11413,PhysRevLett.66.1773,RevModPhys.78.17}.

Over the past few decades, extensive research has been conducted to investigate the phase diagram of the $J_1$-$J_2$ square lattice Heisenberg model. A general agreement has been established that when $J_2 \rightarrow 0$, the model's ground state manifests as a N\'eel antiferromagnetic order \cite{PhysRevB.56.11678}, which extends to a finite region of $J_2/J_1 \approx 0.5$. For very large but finite values of $J_2/J_1$, the model's ground state has antiferromagnetic (AFM) stripe order \cite{PhysRevLett.63.2148,JPSJ.84.024720}. Intriguingly, the intermediate regime of $0.5\lesssim J_2/J_1 \lesssim 0.6$ is termed the non-magnetic phase, which has been a focal point of ongoing research. Many methods have been employed to study this model, including the exact diagonalization \cite{PhysRevB.41.4619,PhysRevLett.63.2148,PhysRevB.43.10970,refId0,Mambrini2006PlaquetteVC}, series expansion \cite{PhysRevB.54.3022,PhysRevB.60.7278,PhysRevB.73.184420}, quantum Monte Carlo \cite{JPSJ.84.024720,PhysRevB.88.060402,PhysRevB.102.014417}, and tensor network methods \cite{PhysRevB.86.024424,PhysRevLett.113.027201,PhysRevLett.121.107202,PhysRevB.97.174408,SciPostPhys.10.1.012,LIU20221034,PhysRevX.11.031034,PhysRevB.96.014414,LIU2024190,PhysRevB.109.235116}. 

Despite the valuable insights gained from these results, which have significantly enhanced our understanding of the phase diagram of the $J_1$-$J_2$ square lattice Heisenberg model, the nature of the non-magnetic regime continues to be a subject of active discussion. A variety of competing states have been suggested for this regime with different methods. These potential states encompass the plaquette valence bond solid (PVBS) \cite{PhysRevLett.84.3173,PhysRevLett.91.197202,Mambrini2006PlaquetteVC,PhysRevB.78.214415,PhysRevB.85.094407,PhysRevB.89.104415,PhysRevLett.113.027201,PhysRevB.102.014417,PhysRevX.11.031034,PhysRevLett.121.107202}, columnar valence bond solid (CVBS) \cite{PhysRevLett.63.2148,PhysRevB.43.10970,Schulz_1992,PhysRevB.97.174408,PhysRevB.60.7278,PhysRevLett.66.1773}, and quantum spin liquid states \cite{PhysRevLett.87.097201,PhysRevLett.91.067201,PhysRevLett.111.037202,PhysRevB.88.060402,JPSJ.84.024720,PhysRevB.96.014414,PhysRevLett.121.107202,PhysRevB.98.241109,PhysRevB.102.014417,PhysRevX.11.031034}. The source of these controversies can be attributed to the difficulty of accurately simulating large-scale quantum many-body systems. 

{Recently, an approach termed Fully Augmented Matrix Product States (FAMPS), has been developed to address and mitigate the limitations inherent in the Density Matrix Renormalization Group (DMRG) method when applied to the study of two-dimensional systems} \cite{Qian_2023}. This method has been subsequently applied to investigate the phase diagram of $J_1$-$J_2$ square lattice Heisenberg model \cite{PhysRevB.109.L161103}. In \cite{PhysRevB.109.L161103}, it is found that a VBS phase directly connects N\'eel and stripe AFM phases, indicating the absence of a spin liquid phase in the $J_1$-$J_2$ square lattice Heisenberg model
(see Fig.~\ref{model} (b)). However, the exact nature of the VBS phase, whether it is a PVBS or a CVBS, remains undetermined. To address this long-standing issue, we employ the DMRG and FAMPS methods, to investigate the nature of the VBS phase at $J_2 = 0.57$,  which is deep in the VBS phase.

By applying an anisotropy $\Delta_y$ in the nearest neighboring coupling in the $y$-direction in the $J_1$-$J_2$ square lattice Heisenberg model, we can detect the possible spontaneous rotational symmetry breaking of the VBS phase, which can help us to distinguish CVBS from PVBS. We push the bond dimension to $D = 25000$ in our FAMPS calculations and simulate systems with a maximum size of $14 \times 14$. Through meticulous extrapolation with truncation errors and reliable finite-size and finite $\Delta_y$ scaling analysis, we establish that the VBS phase in the $J_1$-$J_2$ square lattice Heisenberg model is a PVBS type.

The rest of this paper is structured as follows: in Sec.~\ref{sec:II}, we introduce the model in which we add an anisotropy $\Delta_y$ in the nearest neighboring coupling in the $y$-direction  to the $J_1$-$J_2$ square lattice Heisenberg model. We also outline the methods used to study this model and demonstrate the advantage of the FAMPS method in comparison to the pure DMRG method.
In Sec.~\ref{sec:III}, we present the results of the VBS dimer order parameter and discuss the approach to determine the nature of the VBS phase. We conclude our work in Sec.~\ref{sec:IV}.

\section{Model and Methods}
\label{sec:II}

\subsection{Model}

The Hamiltonian of $J_1$-$J_2$ square lattice Heisenberg model is given by
\begin{equation}
		H = J_1\sum_{\langle i,j\rangle} \hat{S}_i \cdot \hat{S}_j + J_2\sum_{\langle\langle i,j\rangle\rangle} \hat{S}_i \cdot \hat{S}_j
	\label{eq:ham_1}
\end{equation}
with $\langle i,j\rangle$ denoting the nearest neighboring sites and $\langle\langle i,j\rangle\rangle$ denoting the next-nearest neighboring sites. $\hat{S_i}$ is the spin-1/2 operator at site $i$ and $J_1$ and $J_2$ are the nearest and next-nearest neighboring couplings, respectively.

\begin{figure}[t]
	\includegraphics[width=80mm]{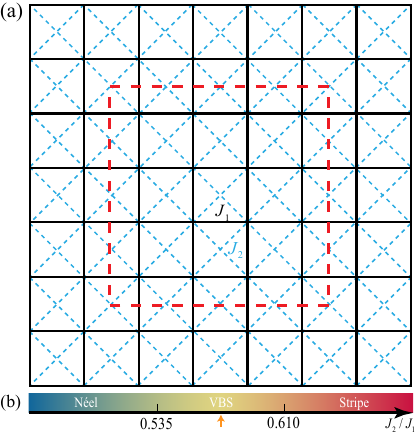}
	\caption{(a) Illustration of an $8 \times 8$ $J_1$-$J_2$ Heisenberg model on a square lattice, where black lines indicate nearest neighboring couplings $J_1$ and dashed blue lines represent next-nearest neighboring couplings $J_2$. Open boundary conditions are applied in both $x$ and $y$ directions. The central zone marked by a red dashed area highlights the lattice region utilized for calculating the VBS dimer order parameter $D_\alpha$ defined in Eq.~(\ref{eq:D_alpha}) (for each system, we use only the results in the central half for the calculation of $D_\alpha$).
	(b) Phase diagram of the $J_1$-$J_2$ square lattice Heisenberg model, illustrating a VBS phase sandwiched by the N\'eel and stripe AFM phases within the coupling ratio range $0.535 \lesssim J_2/J_1 \lesssim 0.610$ \cite{PhysRevB.109.L161103}. The orange arrow at $J_2 = 0.57$ denotes the specific coupling strength at which the VBS dimer order parameter is calculated in this work.}
	\label{model}
\end{figure}

Our study focuses on a square lattice with $L_x = L_y$ with open boundary conditions in both directions, which preserves the rotational symmetry. The ground state of the VBS phase could potentially manifest as a twofold degenerate CVBS or a genuine PVBS. It is known that there is no spontaneous symmetry breaking for systems with finite size, so it is hard to distinguish between these two states directly in finite systems. To address this issue, we introduce an anisotropy $\Delta_y$ in the $y$-direction of the nearest neighboring coupling to detect the possible spontaneous rotational symmetry breaking of the VBS state in the thermodynamic limit. By first extrapolating the system sizes to the thermodynamic limit, followed by the extrapolation to the zero $\Delta_y$ limit, we can determine the true nature of the VBS phase. The new Hamiltonian is given by
\begin{equation}
	\begin{split}
		H = \sum_{r} &J_1 \hat{S}_r \cdot \hat{S}_{r+\hat{x}}+ (J_1 + \Delta_y)\hat{S}_r \cdot \hat{S}_{r+\hat{y}} \\
		&+ J_2 (\hat{S}_r \cdot \hat{S}_{r+\hat{x}+\hat{y}} + \hat{S}_r \cdot \hat{S}_{r+\hat{x}-\hat{y}})
	\end{split}
	\label{eq:ham_2}
\end{equation}
When $\Delta_y$ is set to $0$, this model reduces to the $J_1$-$J_2$ square lattice Heisenberg model. In this work, we set $J_1 = 1.0$ as the energy unit and $J_2=0.57$, deep in the VBS phase as shown in Fig.~\ref{model}(b).

\subsection{Method}

DMRG is a powerful numerical method for studying one-dimensional and quasi-one-dimensional quantum systems \cite{PhysRevLett.69.2863,PhysRevB.48.10345, RevModPhys.77.259}. It is based on the Matrix Product States (MPS) ansatz which is a variational wavefunction ansatz for one-dimensional quantum systems \cite{PhysRevLett.75.3537,PhysRevB.73.094423,MPS}. The MPS ansatz is defined as
\begin{equation}
	|\mathrm{MPS}\rangle = \sum_{\{s_i\}} \text{Tr}(A^{s_1} A^{s_2} \cdots A^{s_{N}}) |s_1 s_2 \cdots s_{N}\rangle
	\label{eq:MPS}
\end{equation}
where $A^{s_i}$ is the rank-3 local tensor at site $i$ with one physical index $s_i$ (with dimension $d$) and two auxiliary indices (with bond dimension $D$). $D$ is the key parameter in DMRG calculations which determines the accuracy of the calculation. Even though DMRG is designed for one-dimensional systems, it can be extended to two-dimensional systems. Nonetheless, the bond dimension required for the accurate simulation of two-dimensional systems increases exponentially with the system size, due to the area-law of entanglement entropy \cite{RevModPhys.82.277}. Hence for DMRG calculations, achieving a sufficiently large bond dimension to study a two-dimensional system with high precision becomes computationally challenging.

Recently, FAMPS method has been introduced to address the limitations of DMRG in two-dimensional systems. This approach enhances the entanglement in MPS and improves computational accuracy by incorporating an additional layer of unitary tensors, known as disentanglers \cite{PhysRevLett.101.110501}, to the physical indices of the MPS. The FAMPS ansatz is defined as
\begin{equation}
	|\mathrm{FAMPS}\rangle = D(u)|\mathrm{MPS}\rangle
	\label{eq:FAMPS}
\end{equation}
where $D(u)$ denotes the additional disentangler layer. FAMPS method has been demonstrated to attain more precise results than pure DMRG method and can support area-law-like entanglement for 2D systems while maintaining the low cost of DMRG $\mathrm{O}(D^3)$ with a small overhead $\mathrm{O}(d^4)$ \cite{Qian_2023,PhysRevLett.126.170603,PhysRevB.105.205102}. 
In this work, we employ the FAMPS method to study the $J_1$-$J_2$ square lattice Heisenberg model at $J_2 = 0.57$ with the largest system size $14\times14$ and largest bond dimension $D = 25000$.
The other system sizes are studied with pure DMRG method with the largest bond dimension $D = 50000$, which gives reliable results with extrapolations with truncation errors in the DMRG calculation.

\begin{figure}
	
	\includegraphics[width=80mm]{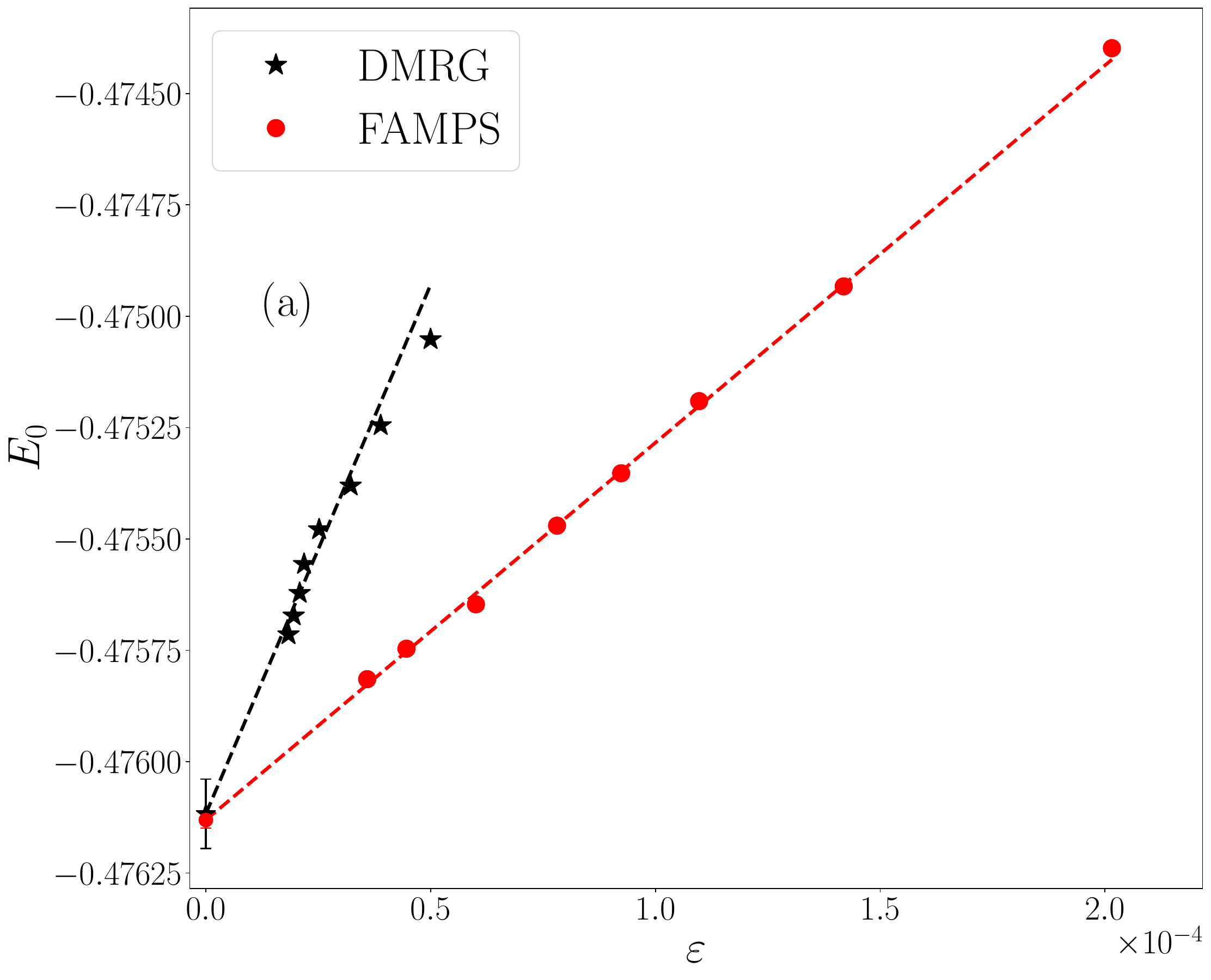}
	\includegraphics[width=80mm]{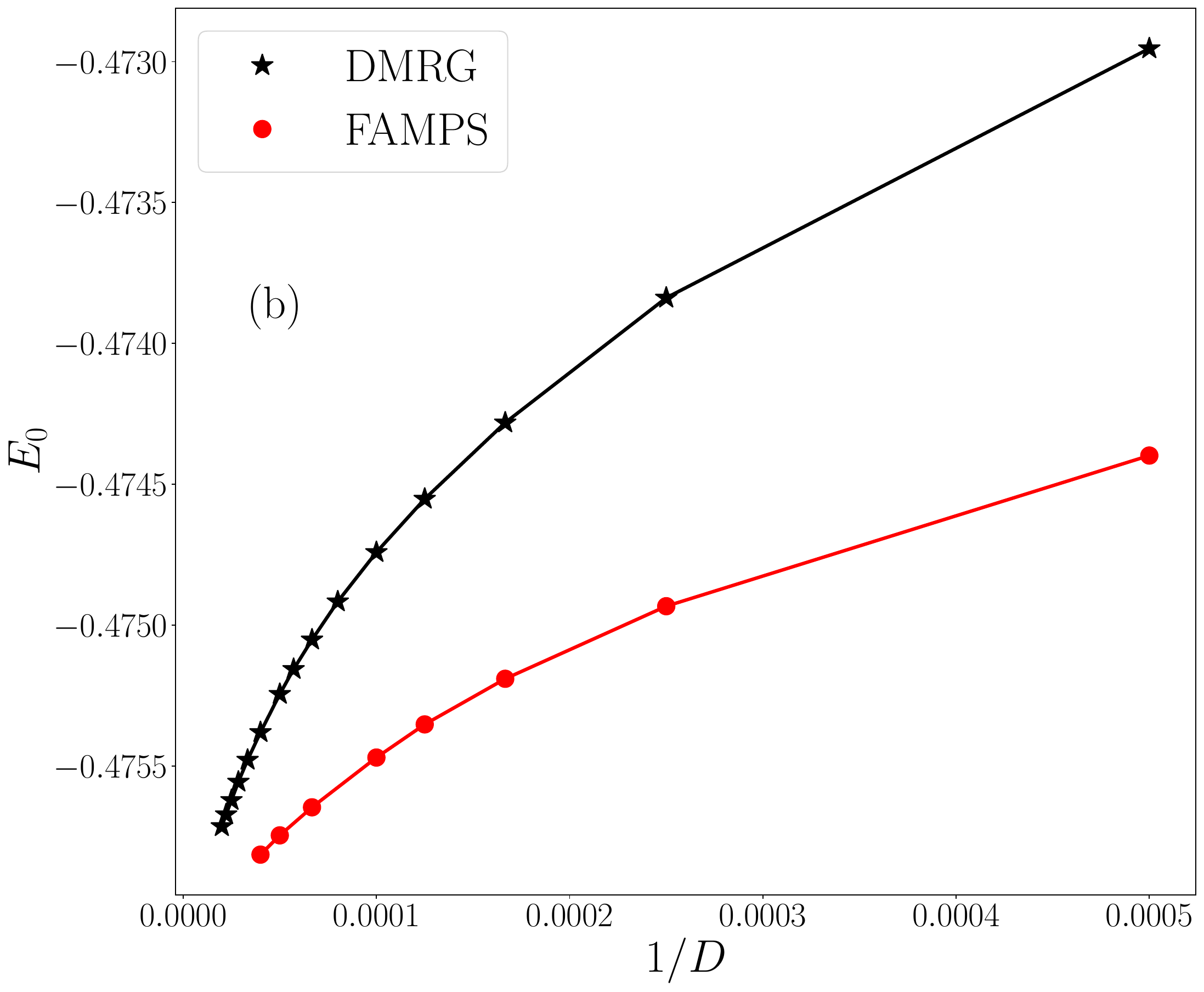}

	\caption{(a) Ground state energy per site $E_0$ for the $J_1$-$J_2$ model with $J_2 = 0.57$, $\Delta_y = 0$ and system size $14 \times 14$ under OBC, as a function of the truncation errors $\varepsilon$. The data is calculated using both FAMPS and pure DMRG methods. The dashed line indicates a linear fitting, which is extrapolated to $\varepsilon = 0$. Though the extrapolated results for DMRG and FAMPS are consistent, the error bar for FAMPS result is quite smaller.
	(b) $E_0$ as a function of the bond dimension $D$. For each $D$, the energy computed using FAMPS is consistently lower than that obtained with pure DMRG method. Analysis indicates that the performance of FAMPS at a given bond dimension $D$ is approximately equivalent to pure DMRG calculation with bond dimension $3D$.
	}
	\label{energy}
\end{figure}

To show the improvement of FAMPS over pure DMRG method {in wider systems}, we calculate the ground state energy per site of the $J_1$-$J_2$ square lattice Heisenberg model (with $\Delta_y = 0$) at $J_2 = 0.57$ for system size $L = 14\times 14$ under open boundary conditions (OBC). The results are presented in Fig.~\ref{energy}. By applying a linear extrapolation to obtain the ground state energy $E_0$ at the truncation errors $\varepsilon = 0$, we observe that both FAMPS and DMRG methods yield consistent results. However, the FAMPS method provides a significantly smaller error bar, indicating a more reliable and accurate estimation of $E_0$ {for the $14 \times 14$ system}. The FAMPS method achieves convergence more rapidly than the pure DMRG approach. Notably, even the data points corresponding to the largest truncation errors in FAMPS {calculations} align with the linear fitting line. In contrast, the data points with large truncation errors from the pure DMRG calculations deviate from the fitting line, indicating less robustness in approaching the ground state {for wider systems} .

The convergence of the energy per site with bond dimension $D$ is depicted in Fig.~\ref{energy} (b). The results highlight that for a given bond dimension, the energy calculated using the FAMPS method consistently falls below that calculated by the pure DMRG method.  Notably, the {simulation accuracy} of FAMPS at a specific bond dimension $D$ is approximately equivalent to that of DMRG with bond dimension $3D$. For instance, the energy value obtained with FAMPS at $D = 15000$ is comparable to those calculated with DMRG at $D = 45000$.

\section{Results}
\label{sec:III}

\begin{figure}[b]
	\includegraphics[width=80mm]{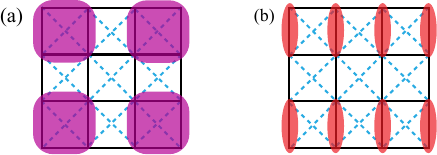}
	\caption{(a) PVBS state with $D_x$ equals to $D_y$. (b) CVBS state with $D_y$ is non-zero and $D_x$ is vanishing.}
	\label{VBS}
\end{figure}

The VBS dimer order parameter, denoted as $D_\alpha$ ($\alpha = x/y$), serves as a critical observable for distinguishing between the competing VBS states in the $J_1$-$J_2$ square lattice Heisenberg model \cite{PhysRevB.98.241109,PhysRevB.101.157101}. This parameter effectively captures the possible rotational symmetry-breaking, thereby providing a quantitative measure to differentiate between the CVBS and PVBS states. The definition of $D_\alpha$ ($\alpha = x/y$) is given by
\begin{equation}
	D_\alpha = \frac{1}{N_b}\sum_r e^{i q_\alpha \cdot r} \hat{S}_r \cdot \hat{S}_{r+\alpha} 
	\label{eq:D_alpha}
\end{equation}
where $N_b$ is the number of bonds in the calculation, $q_\alpha$ is the wave vector associated with the VBS order, taking values $(\pi, 0)$ or $(0, \pi)$ depending on the direction $\alpha$ of the dimer order. In the thermodynamic limit, for the CVBS state, either $D_x$ or $D_y$ will be non-zero while the other approaches zero. Conversely, for the PVBS state, both $D_x$ and $D_y$ should exhibit non-zero values, and the ratio $|\langle D_x \rangle| / |\langle D_y \rangle|$ should be unity, indicating an isotropic dimer distribution across the lattice. The illustration of the PVBS and CVBS dimer is shown in Fig.~\ref{VBS}.

\begin{figure}
	\includegraphics[width=80mm]{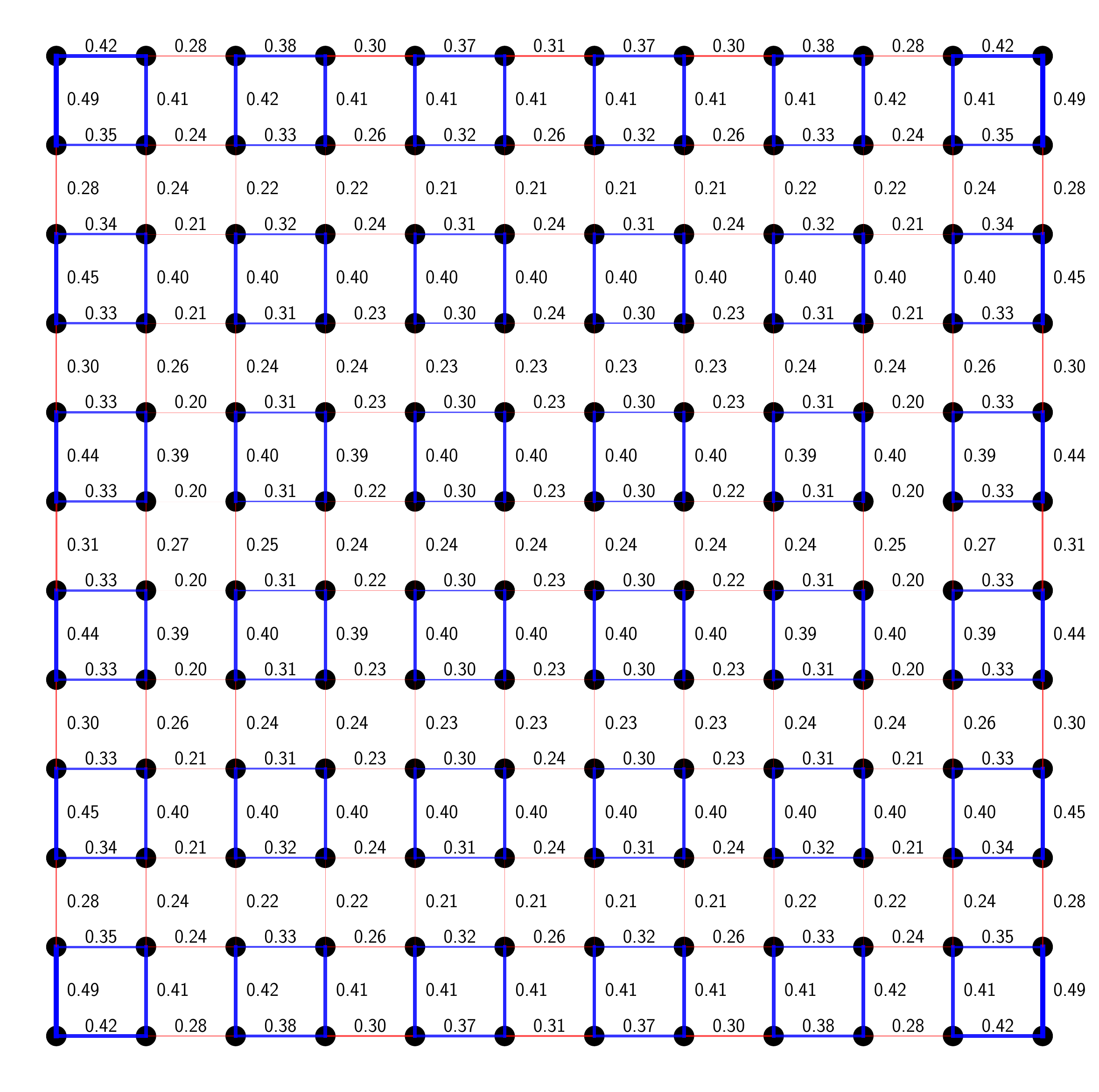}
	\caption{The bond strength $|\langle \hat{S}_r\cdot \hat{S}_{r+\alpha} \rangle|$ of the $12 \times 12$ system  at $\Delta_y = 0.03$. The blue line shows the strong coupling bonds and the red line shows the weak coupling bonds. 
	}
	\label{bond_energy}
\end{figure}

\begin{figure}[t]
	\includegraphics[width=80mm]{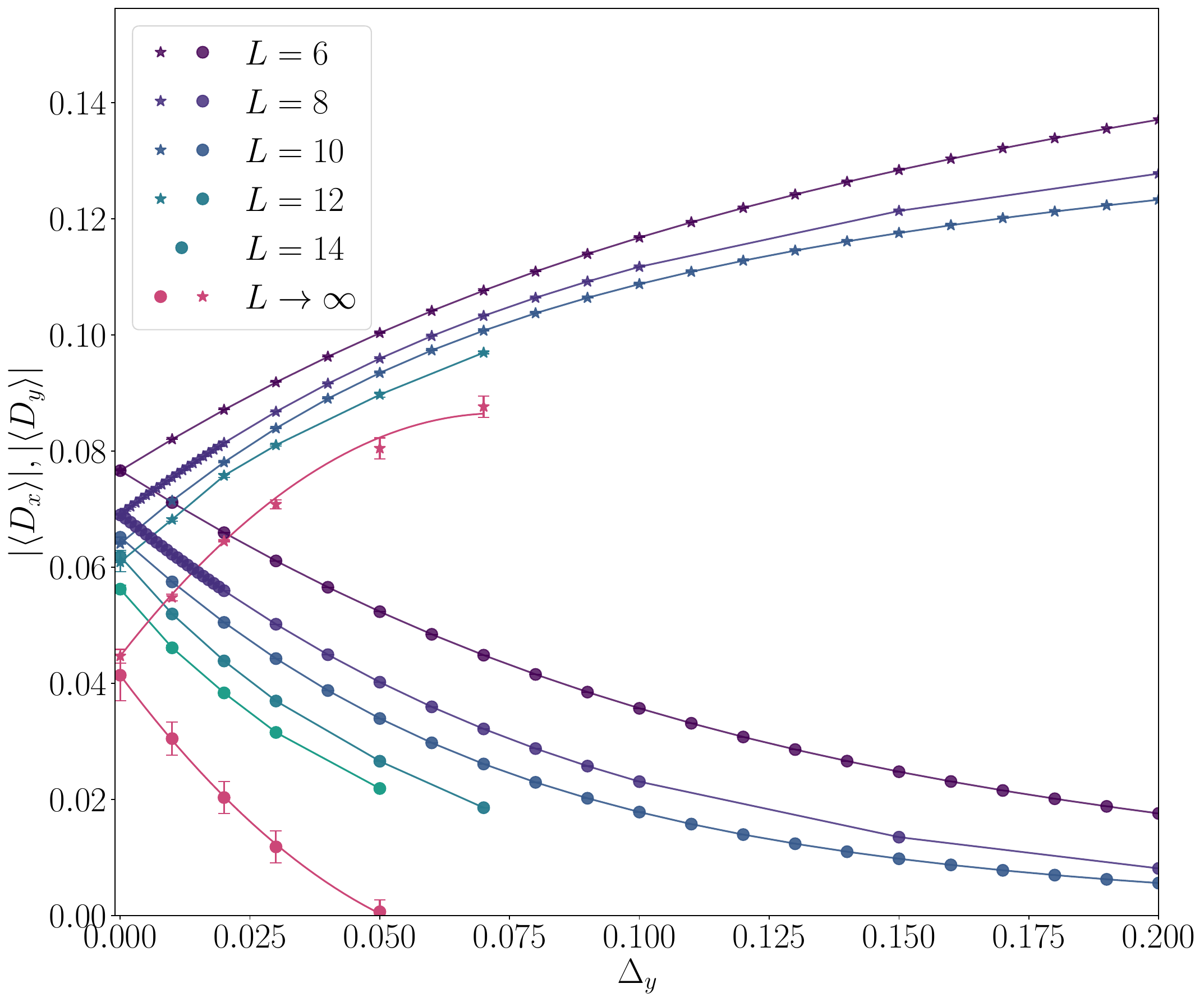}
	\caption{The VBS dimer order parameter $D_\alpha$ versus $\Delta_y$ with $D_y$ denoted as star markers and $D_x$ denoted as circle markers. All data have been extrapolated to truncation errors $\varepsilon = 0$.  The red markers represent values extrapolated to the thermodynamic limit ($L \rightarrow \infty$). Quadratic fitting is applied to these extrapolated points versus $\Delta_y$. The convergence of fitting lines at $\Delta_y = 0$ strongly suggests that there is no spontaneous rotational symmetry breaking in the VBS phase and the VBS phase is a PVBS type.}
	\label{main_reslut}
\end{figure}

In our study, by adding an anisotropy $\Delta_y$ in the nearest neighboring coupling in the $y$-direction to the $J_1$-$J_2$ square lattice Heisenberg model, we can detect the possible spontaneous rotational symmetry breaking of the VBS state. As shown in Fig.~\ref{bond_energy}, though the strong coupling bonds (with larger $|\langle \hat{S}_r\cdot \hat{S}_{r+\alpha} \rangle|$) form a plaquette structure, the rotational symmetry is broken due to the inclusion of the $\Delta_y$ term in Eq.~(\ref{eq:ham_2}). This rotational symmetry breaking can be detected by the VBS dimer order parameter.

In Fig.~\ref{main_reslut}, we present the main results about VBS dimer order parameter $D_\alpha $ as a function of the anisotropy $\Delta_y $ for $J_2 = 0.57 $ across various system sizes $L \times L$. To reduce the finite-size effect, we only use results in the central half of the studied systems. We can find that as $\Delta_y $ increases, $D_x$ ($D_y$) decreases (increases). For $\Delta_y \gtrsim 0.2 $, $D_x $ almost approaches zero while $D_y$ has a finite values. This behavior is consistent with theoretical expectations of Eq.~(\ref{eq:ham_2}), where rotational symmetry is explicitly broken by including the $\Delta_y$ term. We also notice that as expected, when $\Delta_y = 0$, $D_x$ = $D_y$ for given systems, because spontaneous symmetry breaking can't occur for finite systems. To detect the possible spontaneous rotational symmetry breaking, we need to focus on how $D_{\alpha}$ behaves as $\Delta_y$ approaches zero.

To accurately determine the true VBS phase type at $\Delta_y = 0$, we have to extrapolate $D_\alpha$ to the thermodynamic limit $L \rightarrow \infty$. {In this process, the order in which the limit is taken is critical.} 
We first approach the thermodynamic limit and then take the limit of vanishing anisotropy $\Delta_y$. This approach can be formally expressed as
\begin{equation}
	D_\alpha = \lim_{\Delta_y \rightarrow 0} \lim_{L \rightarrow \infty} D_\alpha
	\label{eq:limit}
\end{equation}

In this study, we first perform a linear extrapolation to $L \rightarrow \infty $ for fixed $\Delta_y$. An example for $\Delta_y =0.03$ is shown in Fig.~\ref{finite_size_0.03}, and results for other $\Delta_y$ values are shown in Supplemental Material \cite{sup}. 
We subsequently apply a quadratic fitting of extrapolated $D_\alpha$ with $\Delta_y$ to obtain the spontaneous order parameters. The results are shown in Fig.~\ref{main_reslut}.

\begin{figure}
    \includegraphics[width=80mm]{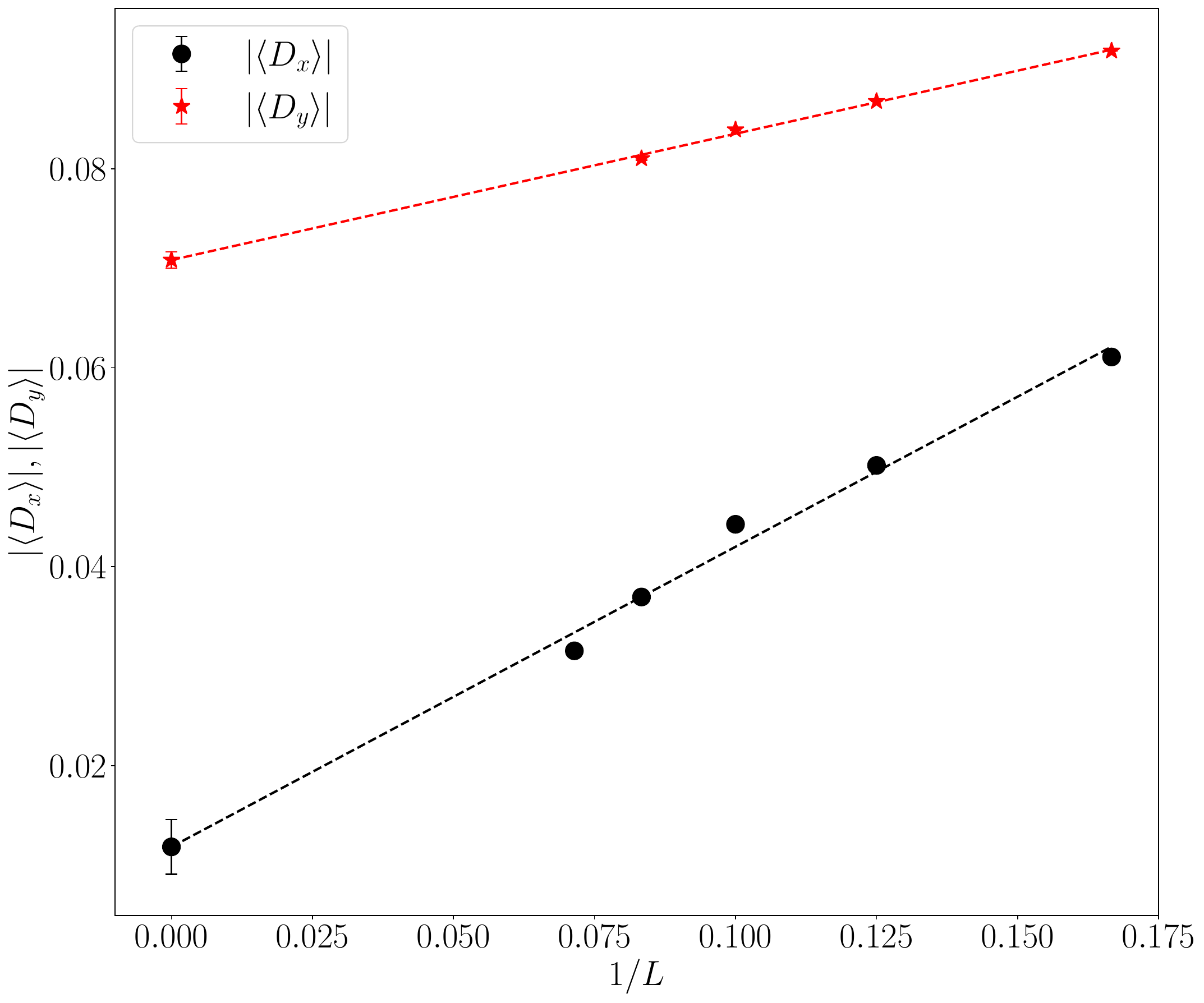}
    \caption{The VBS dimer order parameter $D_\alpha$ versus $1/L$ for $\Delta_y = 0.03$. The dashed line represents the linear fitting to the data points. For $D_x$, we use data points from $6 \times 6$ to $14 \times 14$, while for $D_y$, we exclude the data point from the $14 \times 14$ system because it is not converged in the FAMPS calculation (details can be found in Supplemental Material \cite{sup}). 
	}
	\label{finite_size_0.03}
\end{figure}

If the VBS phase was a CVBS type, $D_x$ would vanish and $D_y$ remain non-zero, due to the applied anisotropy $\Delta_y$ in the $y$-direction of the nearest neighboring coupling. But we find that the fitting curves for $D_x$ and $D_y$ exhibit remarkable convergence at $\Delta_y = 0$, as depicted in Fig.~\ref{main_reslut}. This convergence strongly suggests that the VBS phase in the $J_1$-$J_2$ square lattice Heisenberg model is a PVBS type.

One may notice that in the extrapolation process to $L \rightarrow \infty$, we utilize data points from the $14 \times 14$ system size for $D_x$ but exclude them for $D_y$ (see Fig.~\ref{finite_size_0.03}). The reason for this procedure is as follows. The $D_y$ values obtained from the $14 \times 14$ system appear unreasonably high. Moreover, the linear and quadratic extrapolations of $D_y$ with truncation errors in FAMPS give inconsistent values, suggesting a significant finite-bond effect for $D_y$ in the FAMPS calculation. However, the behavior of $D_x$ with truncation errors in FAMPS is quite flat and the extrapolation with truncation errors is robust. More details of the finite-bond effect are discussed in Supplemental Material \cite{sup}.

In summary, our results indicate that the VBS phase in the $J_1$-$J_2$ square lattice Heisenberg model is a PVBS type.

\section{Conclusion}
\label{sec:IV}
In this work, we employ FAMPS and DMRG methods to the $J_1$-$J_2$ square lattice Heisenberg model to investigate the nature of the VBS phase in its phase diagram. By introducing an anisotropy parameter $\Delta_y$ in the nearest neighboring coupling in the $y$-direction into the system, we are able to distinguish between the CVBS and PVBS types. Through careful extrapolation of truncation errors and employing reliable finite-size scaling, followed by finite $\Delta_y$ scaling analysis of the VBS dimer order parameter, we establish that the VBS phase in the $J_1$-$J_2$ square lattice Heisenberg model is a PVBS type, meaning the rotational symmetry is not spontaneously broken in the VBS phase.

Our results not only resolve the long-standing issue on the characterization of the VBS phase in the phase diagram of the $J_1$-$J_2$ square lattice Heisenberg model, but also demonstrate the potential of FAMPS to be applied to other complex two-dimensional systems, where large system sizes and bond dimensions are necessary to capture the intricate details of the ground state.

\begin{acknowledgments}
We thank useful discussion with Jong Yeon Lee. The calculation in this work is carried out with TensorKit \cite{foot1}.
We acknowledge the support from the Innovation Program for Quantum Science and Technology (2021ZD0301902),
the National Natural Science Foundation of China (Grant
No. 12274290) and the sponsorship from Yangyang Development Fund.

\end{acknowledgments}

\bibliography{reference}

\appendix

\section{Finite-size scaling}
\begin{figure*}
	\includegraphics[width=80mm]{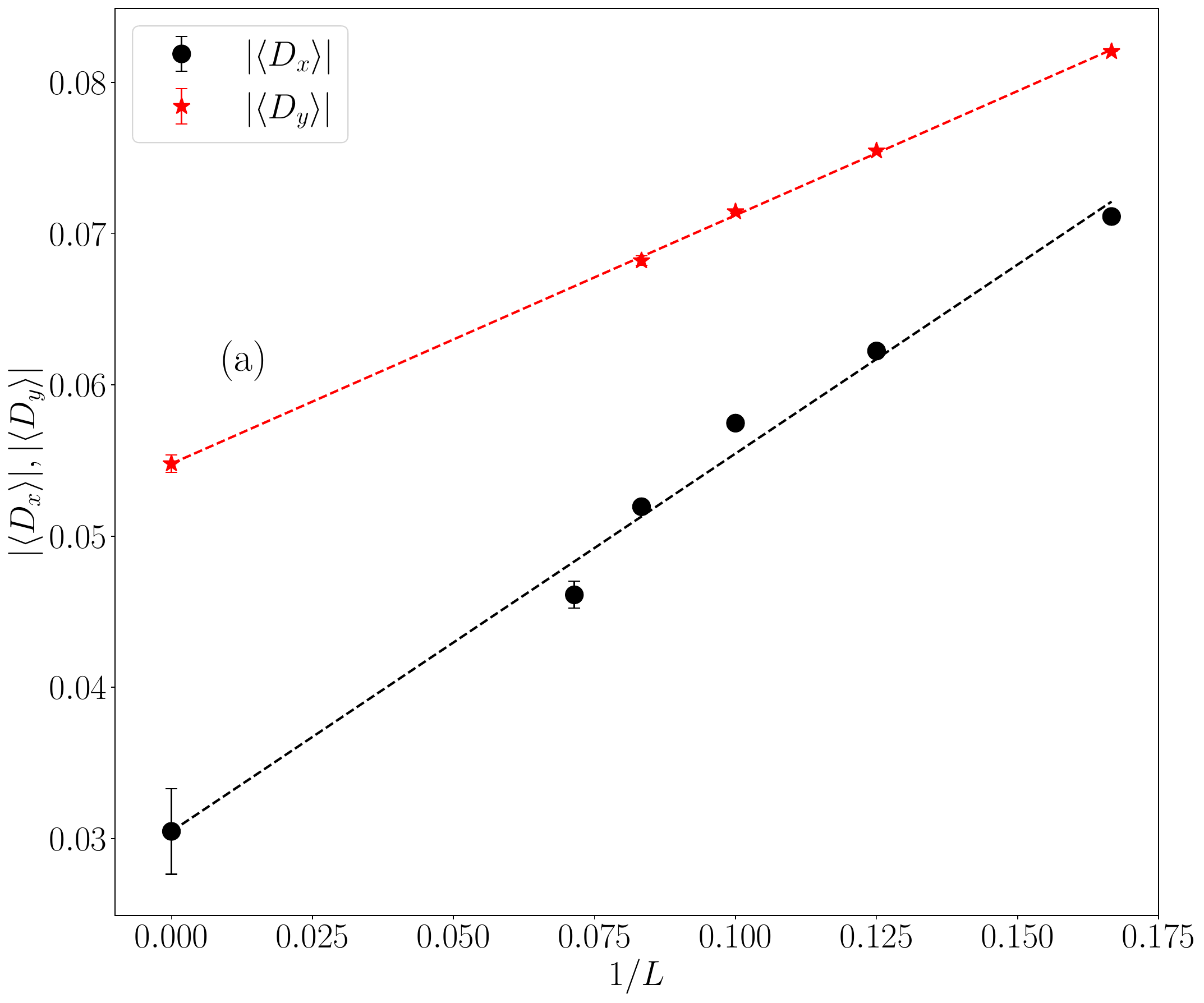}
	\includegraphics[width=80mm]{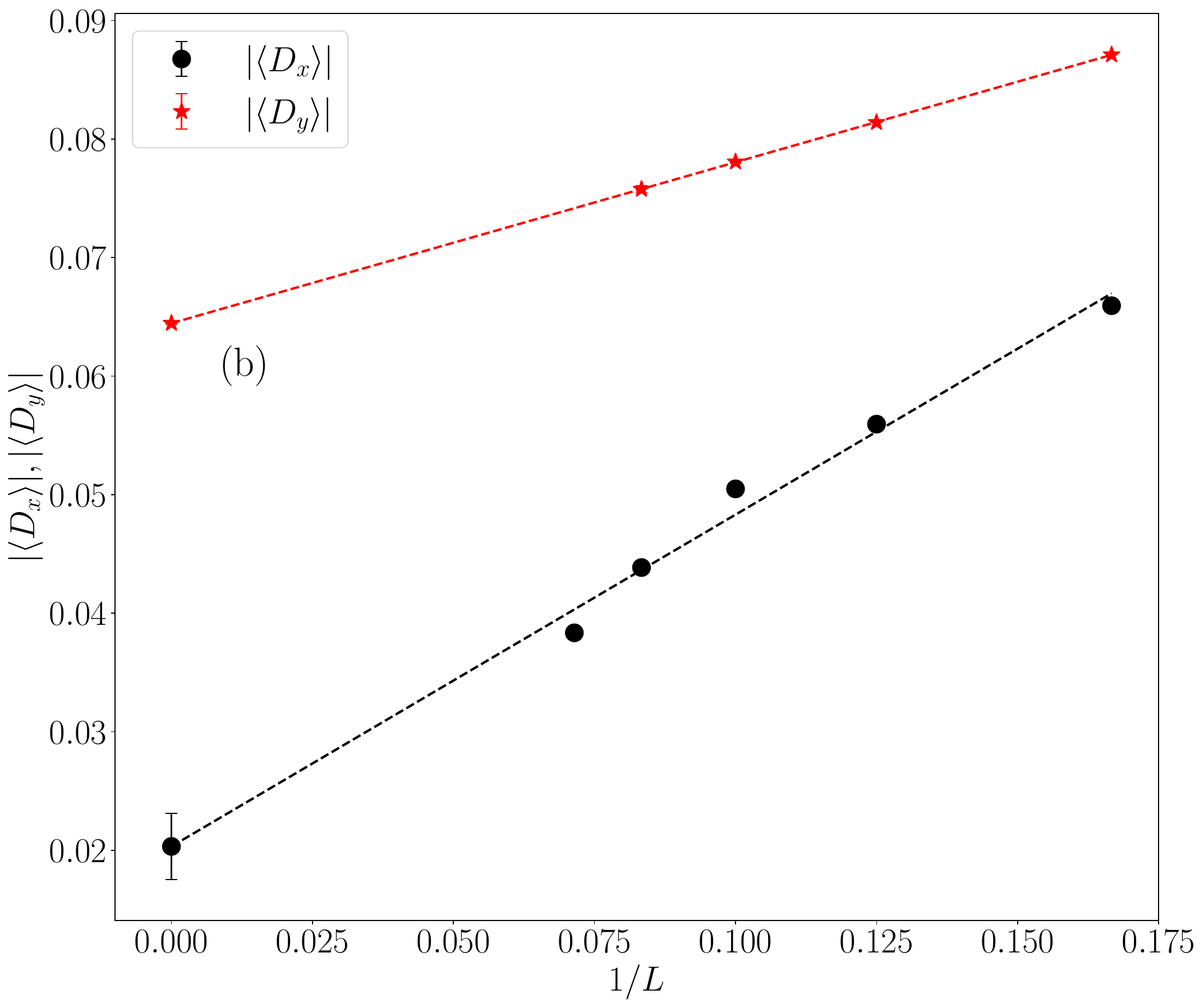}
	\includegraphics[width=80mm]{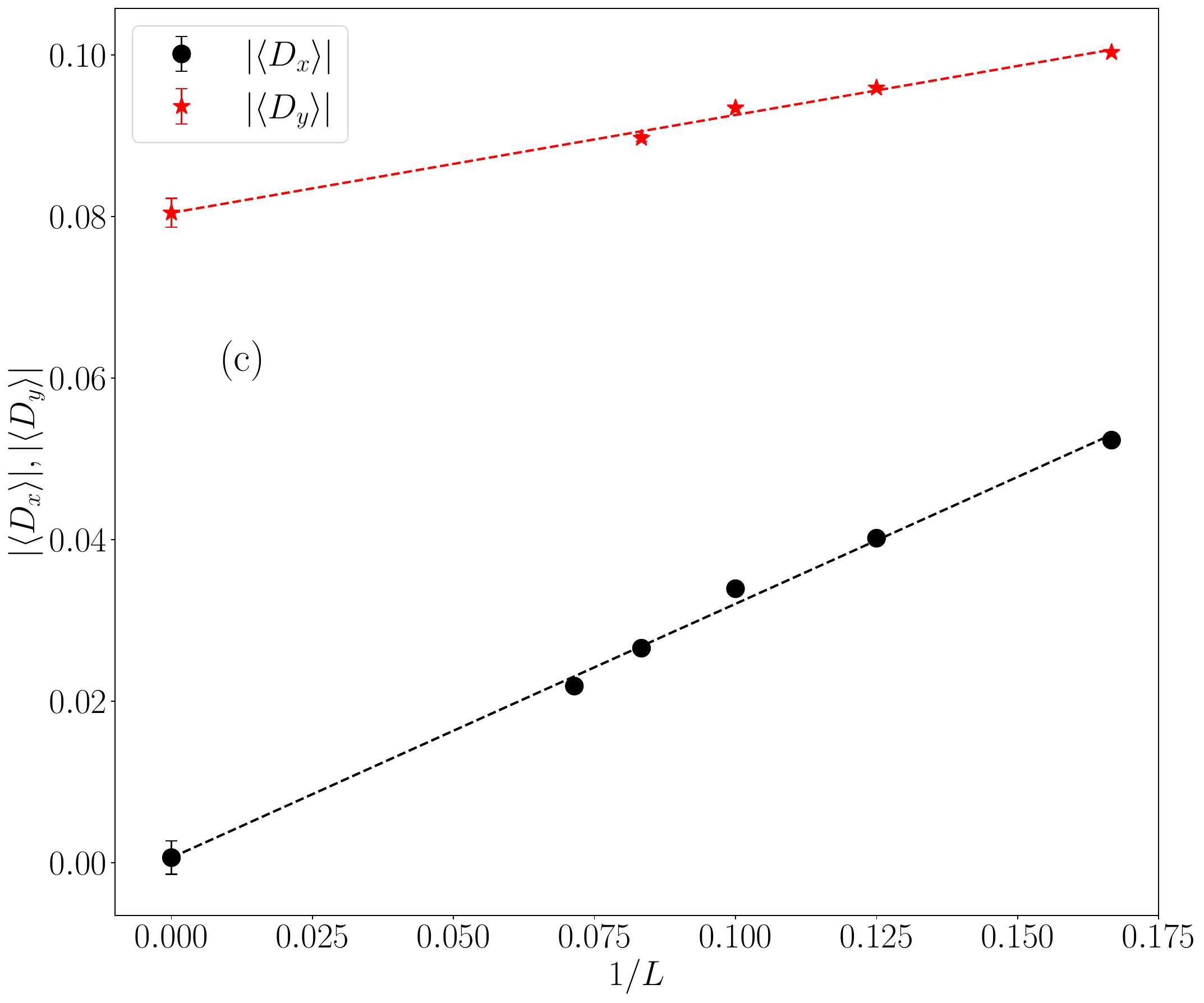}
	\includegraphics[width=80mm]{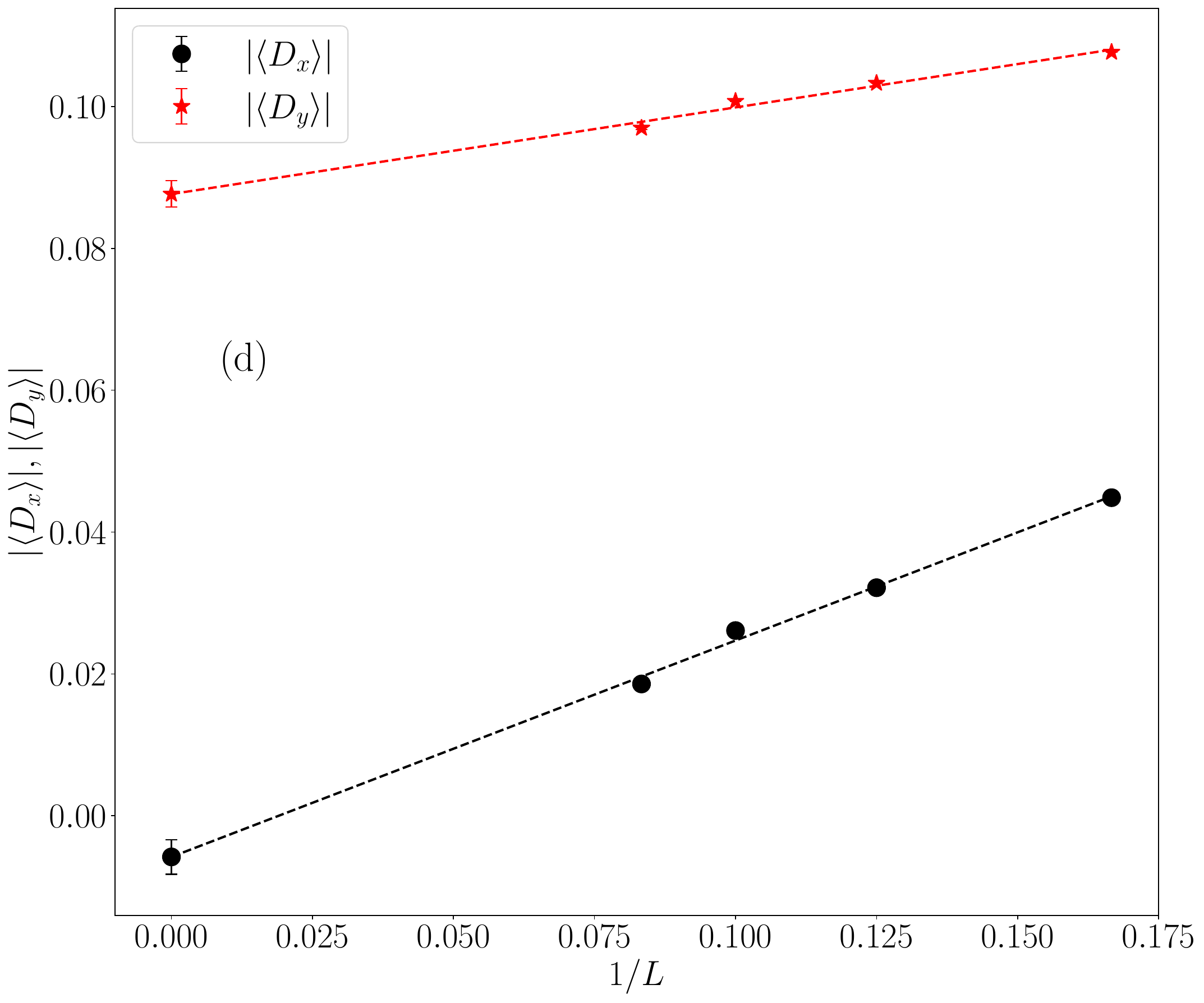}
	\caption{VBS dimer order parameter $D_\alpha$ plotted as a function of system size $1/L$. The dashed line represents a linear fitting to the data points.
	(a) results for $\Delta_y = 0.01$, (b) results for $\Delta_y = 0.02$, (c) results for $\Delta_y = 0.05$, and (d) results for $\Delta_y = 0.07$.
	}
	\label{sup:finite_size}
\end{figure*}

In Fig.~\ref{sup:finite_size}, we show the extrapolation of the VBS dimer order parameter with respect to the system sizes at $\Delta_y = 0.01, 0.02, 0.05$, and $0.07$. We notice that the extrapolated value of $D_x$ for $\Delta_y = 0.07$ becomes negative. But as $\Delta_y = 0.07$ is close to the boundary where $D_x$ vanishes, the uncertainty in the extrapolation should be large. We also notice that the value of  $D_x$ at $\Delta_y = 0.07$ is actually $0$ within a few error bars. We also exclude it in the extrapolation of $\Delta_y$ in the main text.

\section{Finite-bond effect of FAMPS results for the $14 \times 14$ system}

\begin{figure*}
	\includegraphics[width=80mm]{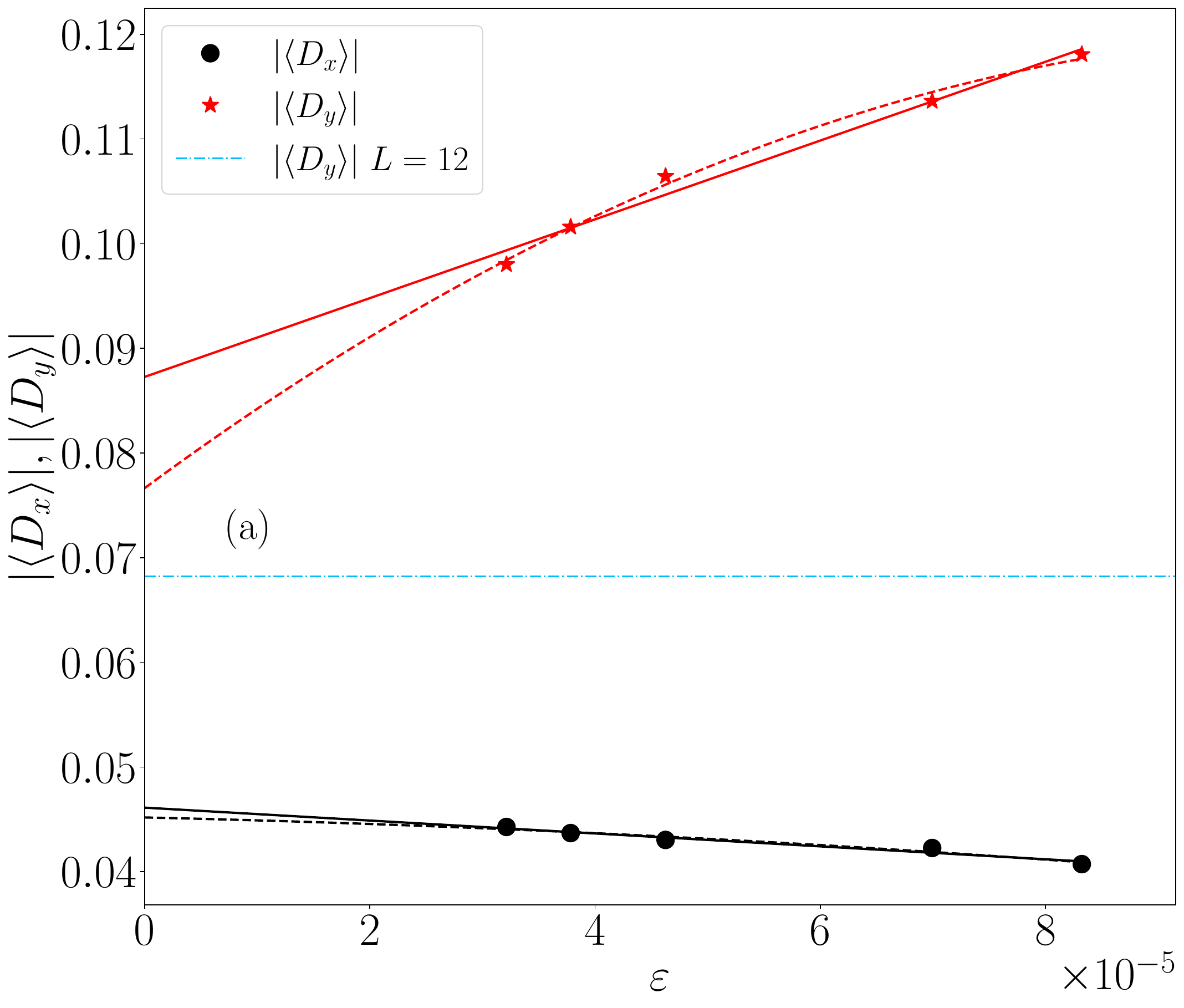}
	\includegraphics[width=80mm]{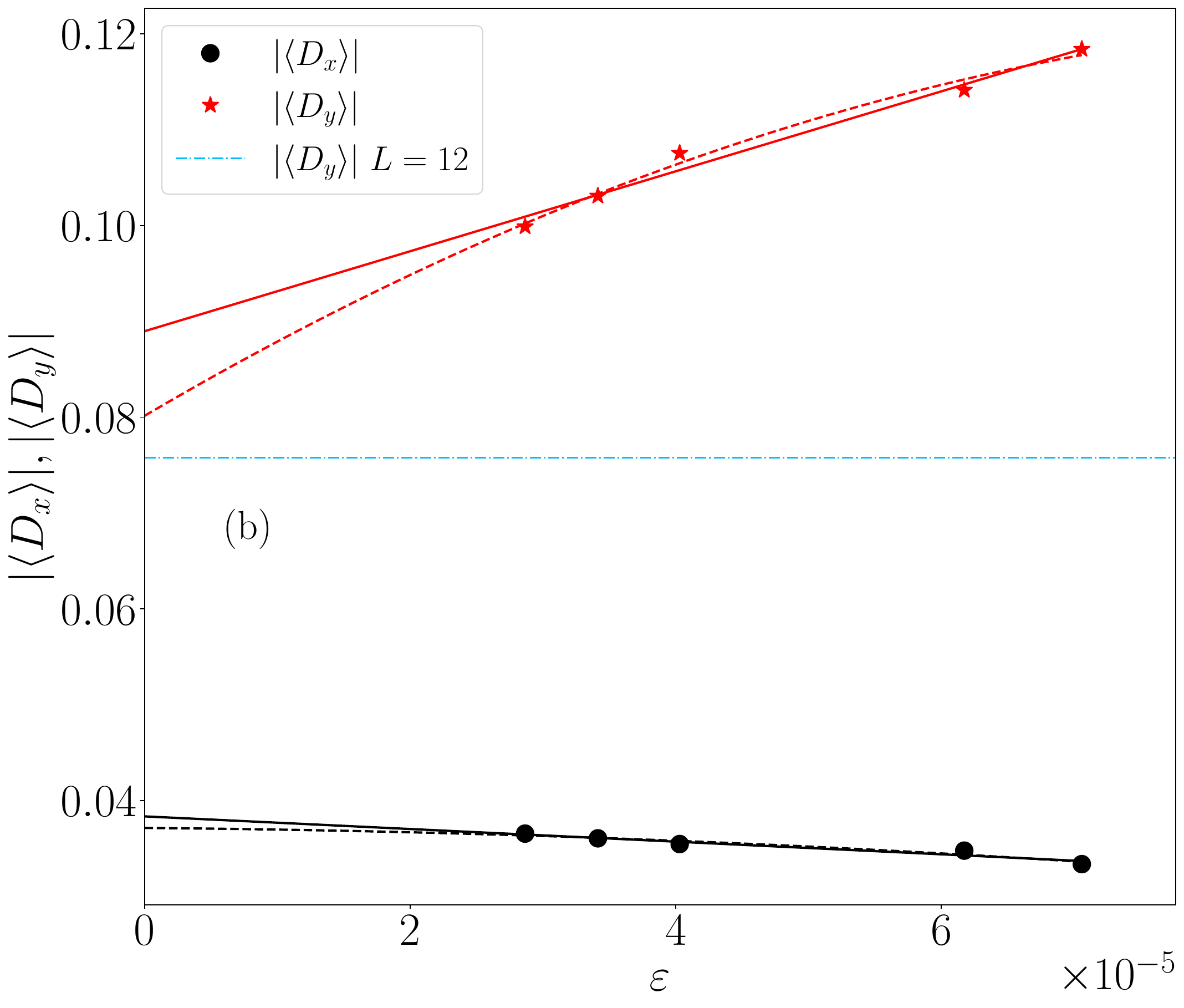}
	\includegraphics[width=80mm]{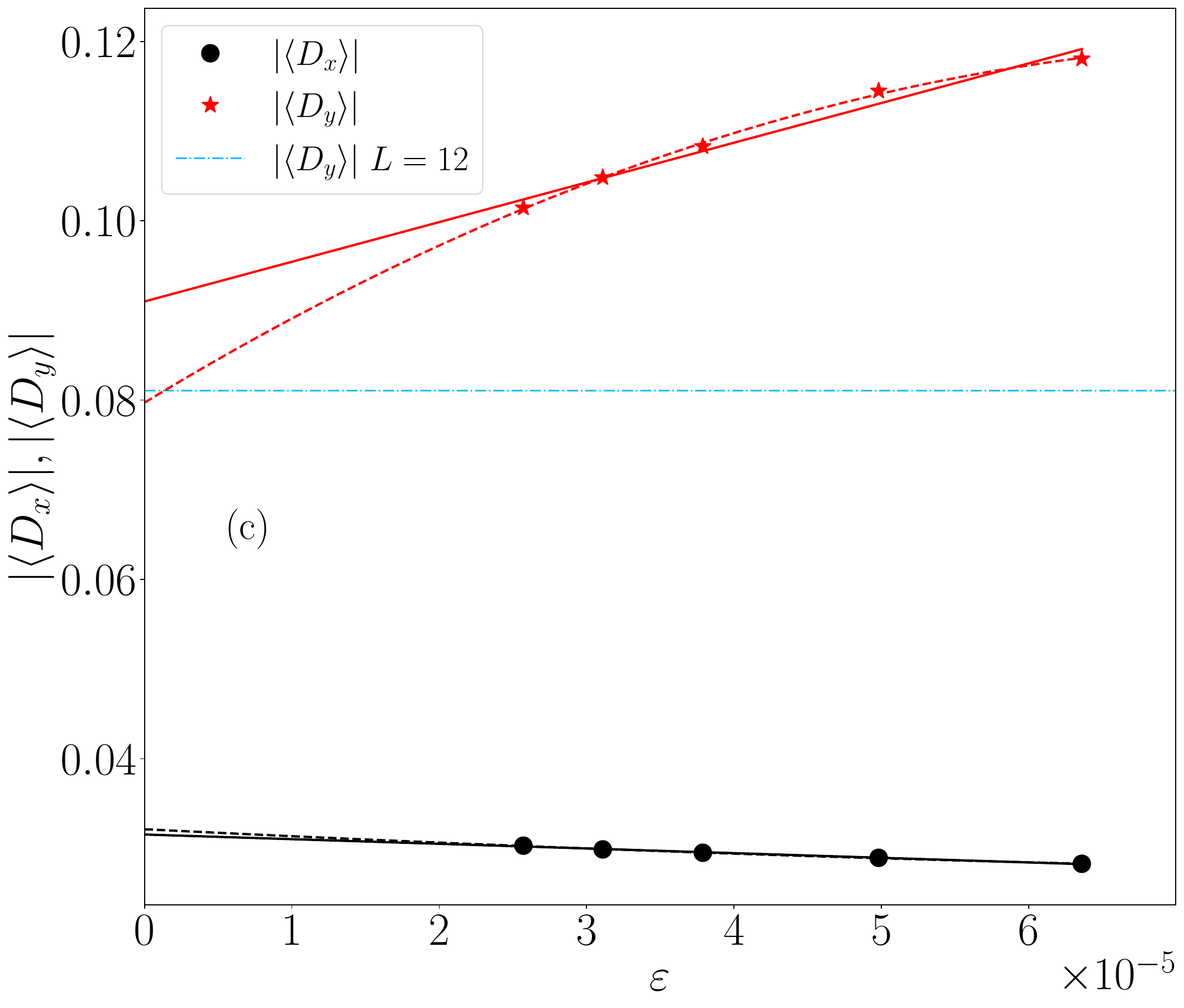}
    \includegraphics[width=80mm]{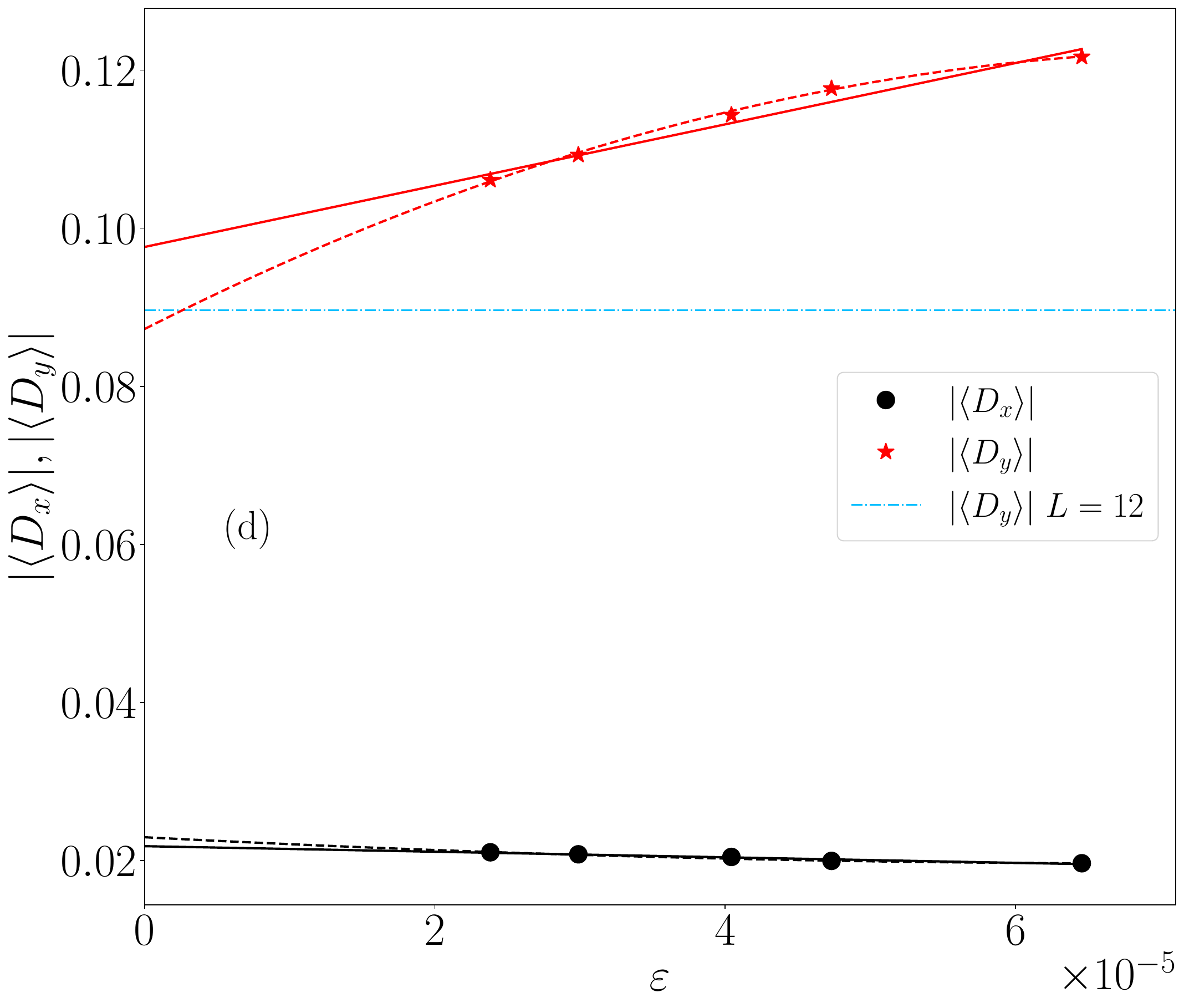}
	\caption{VBS dimer order parameter $D_\alpha$ plotted as a function of truncation errors $\varepsilon$ for the $14 \times 14$ system. The solid line indicates a linear fitting to the data points, while the dashed line represents a quadratic fitting. The blue dot-dash line represents the extrapolated value of $D_y$ for the $12 \times 12$ system as a reference. (a) $\Delta_y = 0.01$, (b) $\Delta_y = 0.02$, (c)$\Delta_y = 0.03$ and (d) $\Delta_y = 0.05$. We can find that the extrapolation for $D_x$ is robust, whereas it is not reliable for $D_y$.}
	\label{sup:finite_bond}
\end{figure*}

To elucidate the finite-bond effect on the VBS dimer order parameter, particularly the unreliability of the results for $D_y$ at system size $14 \times 14$, we present the extrapolation of the VBS dimer order parameter with respect to the truncation {errors} in FAMPS calculations for the system with size $14 \times 14$  in Fig.~\ref{sup:finite_bond}. For both $D_x$ and $D_y$, linear and quadratic fittings are employed to extrapolate the truncation errors to $\varepsilon = 0$. Notably, both fittings give consistent results for $D_x$. However, for $D_y$, linear and quadratic fittings give different results.

Due to the computational resource limitation, achieving the required bond dimension for the convergence of $D_y$ in the $14 \times 14$ system size is challenging. So we exclude the data for size $14 \times 14$ when performing the finite-size scalings for $D_y$.

\end{document}